\documentclass[amsmath,superscriptaddress,showpacs,pre]{revtex4}
\usepackage{graphicx}
\pdfoutput=1
\usepackage{amsmath}
\usepackage{epsfig}
\usepackage{graphics}
\usepackage{color}
\usepackage{latexsym}
\usepackage{amsfonts}
\usepackage{amssymb}
 \usepackage{color}  
\usepackage{plain}   

\begin{document}

\title{Entropy production  and fluctuation theorems under feedback control: the molecular refrigerator model revisited}
\author{T. Munakata}
\affiliation{Department of Applied Mathematics and Physics, 
Graduate School of Informatics, Kyoto University, Kyoto 606-8501, Japan
} 
\email{tmmm3rtk@hb.tp1.jp}
\author{M.L. Rosinberg}
\affiliation{Laboratoire de Physique Th\'eorique de la Mati\`ere Condens\'ee, Universit\'e Pierre et Marie Curie\\ 4 place Jussieu, 75252 Paris Cedex 05, France}
\email{mlr@lptmc.jussieu.fr}

\begin{abstract}
We revisit the model of a Brownian particle in a heat bath submitted to an actively controlled force proportional to the velocity that leads to thermal noise reduction ({\it cold damping}). We investigate the influence of the continuous feedback on the fluctuations of the total entropy production and show that the explicit expression of the detailed fluctuation theorem involves different dynamics and observables in the forward and backward processes. As an illustration, we study the analytically solvable case of a harmonic oscillator and calculate the characteristic function of the entropy production in a nonequilibrium steady state. We then determine  the corresponding large deviation function which results from an unusual interplay between `boundary' and `bulk' contributions.
\end{abstract} 


\maketitle

\section{Introduction}

In parallel with the recent developments in nanotechnology and single-molecule manipulations, there is an increasing interest in understanding the stochastic energetics of small systems driven away from thermal equilibrium. In this context, fluctuation theorems (FT) play a central role as they describe exact symmetry properties of the probability distributions of various thermodynamic quantities such as work, heat, or entropy  (see Refs. \cite{K2007,HS2007,S2008,J2011} for recent reviews). Typically, a (detailed) FT is a relation of the form 
\begin{align}
\label{EqFT0}
\frac{P({\cal S}_t=S)}{P({\cal S}_t=-S)}=e^{\zeta S}
\end{align}
where ${\cal S}_t$ is an observable integrated over a time interval $t$ and $\zeta$ is a positive constant. In other words, a FT states that positive fluctuations are exponentially more probable than negative ones, which can be generally ascribed to the breaking of time-reversal symmetry at the level of stochastic trajectories.

Recently, the extension of stochastic thermodynamics and fluctuation theorems to systems under feedback control has become an active field of research (see Ref. \cite{SU2012} and references therein). Feedback loops, in which some microscopic information about the state of the system is used to manipulate its evolution, are indeed important in many engineering applications and also play a crucial role in biological motors. Feedback control may be interpreted as a kind of  ``Maxwell's demon" \cite{TL2000},  which requires to generalize the second law of thermodynamics and to modify the various fluctuation theorems. 

One of the simplest example of  a small classical system under a continuous feedback control is a Brownian ``particle''  in a heat bath submitted to a velocity-dependent external force. This results in a reduction of thermal fluctuations, as illustrated for instance by experiments with an atomic force microscope (AFM)\cite{LMC2000,JTCC2007}. This technique is named ``cold damping" and is now used in a wide variety of optomechanical or electromechanical systems (see e.g. Ref.\cite{KV2008} for a review on optomechanical cooling and Refs.\cite{V2008,B2009} for an application to a gravitational wave detector). A theoretical description of such a  (classical) molecular refrigerator was provided in Refs. \cite{KQ2004,KQ2007} where it was shown that the contraction of  phase space induced by the additional viscous damping force could be interpreted as an {\it entropy pumping} mechanism. In particular, it was claimed in Ref. \cite{KQ2007} that the FT takes the form of Eq. (\ref{EqFT0}) (with $\zeta=1$) when the entropy pumping  term is included in the overall entropy production. The goal of the present study is to reexamine this statement within the path-integral formalism of Langevin dynamics and to specify the dynamics and observables that are associated to the probabilities appearing in the numerator and denominator of Eq. (\ref{EqFT0}). There is indeed a subtlety in the definition of the so-called ``backward'' (or time-reversed) trajectory in a system with velocity-dependent feedback, which makes the observables measured during the forward and backward processes different. Although the issue of time-reversal symmetry of the feedback force was careful discussed in Ref.\cite{KQ2004}, this point was not clearly presented in Ref. \cite{KQ2007} and not fully appreciated in the subsequent literature. Here, for the sake of simplicity, we mainly consider the case of a time-independent nonequilibrium steady state (NESS). We also assume that the measurement process is error free, in contrast with most recent works on feedback control which investigate the properties of the mutual information acquired through a discrete series of  measurements\cite{SU2012,note0}. Our discussion is illustrated by the paradigmatic (but experimentally relevant) example of a harmonic oscillator for which calculations can be performed analytically.

The paper is organized as follows. In section 2, we briefly recall the analysis performed in Refs. \cite{KQ2004,KQ2007} and  we (re)-derive the integral and detailed fluctuation theorems for the entropy production in the specific case of a feedback control  proportional to the velocity, which corresponds to the actual experimental situation\cite{LMC2000,JTCC2007,V2008,B2009}. The stochastic harmonic oscillator is studied in section 3 where we exactly determine the characteristic function of the entropy production in a steady state and investigate the properties of the large deviation function (the complete asymptotic form of the probability distribution function is given in appendix). We  conclude in section 4. 

\section{Entropy production and fluctuation theorems}

We consider a single Brownian particle (or ``system'') in contact with a heat bath at temperature $T$ whose dynamics is governed by the one-dimensional underdamped Langevin equation\cite{KQ2007}
\begin{align}
\label{EqL1}
m\dot v_s=-\gamma v_s-\partial_{x_s} V_{\alpha}(x_s) +g(v_s) +\xi_s
\end{align}
where $v_s\equiv \dot x_s$ is the velocity of the particle at time $s$, $\gamma$ is the friction coefficient and $\xi_s$ is a delta-correlated white noise with variance $2\gamma T$ (Boltzmann's  constant is set to unity throughout this work). $V_{\alpha}(x)$ is a potential that can be externally controlled via a time-dependent parameter $\alpha(t)$, and $g(v)$ is a velocity-dependent force that results from a feedback mechanism which detects the motion of the particle in real time, like in the AFM experiments described in Refs. \cite{LMC2000,JTCC2007}. The force $g(v)$ is a source of entropy production and at constant $\alpha$  the system eventually reaches a NESS where heat is permanently dissipated. The probability distribution  $p(x,v,t)$ of the system in phase space is solution of the corresponding Kramers equation 
\begin{align}
\label{EqKramers}
\partial_t p(x,v,t)=-v\partial_x p(x,v,t) +\frac{1}{m} \partial_v\left\{[\gamma v +\partial_x V_{\alpha}(x)-g(v)] p(x,v,t)+ \gamma \frac{T}{m} \partial_v p(x,v,t)\right\} \ .
\end{align}

Let us first briefly recall the analysis of Refs. \cite{KQ2004,KQ2007} for the entropy production along a single trajectory $\{x_s\}_{s\in[0,t]}$ of the system during a time interval $0\le s\le t$.   Within the stochastic energetics (or thermodynamics) framework\cite{S2008,S1998}, one usually identifies two contributions to the entropy production: 

i) the entropy change in the medium, which corresponds to the heat dissipated in the environment,
\begin{align}
\label{EqSm}
\Delta S_m[\{x_s\}]&\equiv \frac{1}{T}Q[\{x_s\}]=\frac{1}{T}\int _0^t ds\:\dot x_s[\gamma \dot x_s-\xi_s]\nonumber\\
&=-\frac{1}{T}\int _0^t ds\:\dot x_s[m\ddot x_s+\partial_{x_s} V_{\alpha}(x_s)-g(v_s)] \ ,
\end{align}
where  the sign of $Q[\{x_s\}]$ is  here chosen to be positive if the heat flows out of the system into the heat bath, 

ii) the entropy change in the system itself\cite{C1999,S2005} 
\begin{align}
\label{EqS}
\Delta S=-\ln p(x_t,v_t,\alpha_t)+\ln p(x_0,v_0,\alpha_0) \ ,
\end{align}
where the probability distribution, solution of Eq. (\ref{EqKramers}), is evaluated along the stochastic trajectory. (Note that all products of stochastic quantities as in Eq. (\ref{EqSm}) are defined with the Stratonovich prescription\cite{S1998}.) Using the Kramers equation,  one can show that a third contribution  appears in the presence of the velocity-dependent  force $g(v)$,
\begin{align}
\label{EqSpu}
\Delta S_{pu}[\{x_s\}]=\frac {1}{m}\int_0^t ds\: \partial _{v_s} g(v_s) \ ,
\end{align}
which is interpreted  in Refs. \cite{KQ2004,KQ2007} as an ``entropy pumping'' performed by the external agent that manipulates the feedback force (this may be for instance an optical or electromechanical device). In the case of a friction-like control, this contribution is negative (see below Eq. (\ref{Eqpu1})). All these quantities fluctuate from one trajectory to another and only the combination
\begin{align}
\label{EqSigma}
\Sigma[\{x_s\}]\equiv\Delta S_m[\{x_s\}]+\Delta S-\Delta S_{pu}[\{x_s\}]
\end{align}
is always non-negative when performing the ensemble average\cite{KQ2004}. $\Sigma[\{x_s\}]$ can thus be interpreted as the overall entropy production in the ``super-system'' composed of the  particle, the heat bath and the external agent. On the other hand, the ensemble average of $\Delta S_{tot}\equiv \Delta S_m+\Delta S$, the entropy production in the particle (or system) and the bath, can be negative. The velocity-dependent feedback thus implies a modification of the second law of thermodynamics. 

In order to discuss the fluctuation theorems in a more specific framework, we now assume that $g(v)$ is proportional (but opposite) to the particle velocity, i.e. $g(v)=-\gamma' v$ (with $\gamma'>0$), like in the AFM setup described in \cite{LMC2000,JTCC2007}. Thermal fluctuations are reduced by this additional friction force and the effective temperature of the system in a steady state, defined by its average kinetic energy, becomes lower than the heat bath temperature (hence heat permanently flows from the bath to the system).  Specifically, in the case of a harmonic potential, one has $T_{eff}\equiv m<v^2>=T\:\gamma/(\gamma+\gamma') $\cite{LMC2000,JTCC2007,V2008} (see also section 3 below). For a linear feedback, the entropy pumping contribution defined by Eq. (\ref{EqSpu}) does not depend on the stochastic trajectory and it decreases linearly with the observation time
\begin{align}
\label{Eqpu1}
\Delta S_{pu}=-\frac{\gamma'}{m}t \ .
\end{align}

It is well known that one can relate $\Delta S_m[\{x_s\}]$, the entropy change in the medium, to the ratio of the probability functionals for the forward and backward (i.e. time-reversed) trajectories\cite{C1999}. Taking into account the presence of the additional friction coefficient $\gamma'$, the probability of the path $\{x_s\}_{s\in[0,t]}$, given that the system started in  the state $(x_0,v_0)$, has the following expression
\begin{align}
\label{EqProba}
{\cal P}[\{x_s\}\vert x_0,v_0]=C \exp\left[\frac{\gamma+\gamma'}{2m} t-\frac{1}{4\gamma T}\int_0^t ds\: \Big( m\ddot x_s+(\gamma+\gamma' )\dot x_s+\partial_xV_{\alpha}(x_s)\Big)^2\right] 
\end{align}
where $C$ is a normalization factor (see Ref. \cite{IP2006} for an explicit derivation of the path probability associated to an underdamped Langevin equation with a general non-conservative force). In the present case,  it is crucial  not to include the linear term $(\gamma+\gamma')t/(2m)$ appearing in the exponential into the normalization factor. Indeed, since the probability of the  path $\{\hat x(s)\}_{s\in[0,t]}$ defined by the time-reversal operation $\hat x_s\equiv x_{t-s},\hat v_s\equiv -v_{t-s},\hat \alpha_s\equiv \alpha_{t-s}$, is given by
\begin{align}
{\cal P}[\{\hat x_s\}\vert \hat x_0,\hat v_0]=C \exp\left[\frac{\gamma+\gamma'}{2m} t-\frac{1}{4\gamma T}\int_0^t ds\: \Big( m\ddot x_s-(\gamma+\gamma' )\dot x_s+\partial_xV_{\alpha}(x_s)\Big)^2\right] \ ,
\end{align}
where $(\hat x_0,\hat v_0)=(x_t,-v_t)$, one must also change the sign of $\gamma'$  in order to extract $\Delta S_m[x(s)]$ from the ratio of the two probabilities.  This yields
\begin{align}
\label{Eqratio}
\frac{{\cal P}_+[\{x_s\}\vert x_0,v_0]}{{\cal P}_-[\{\hat x_s\}\vert \hat x_0,\hat v_0]}&=\exp\left[\frac{\gamma'}{m} t-\frac{1}{T}\int_0^t ds \:\dot x_s\Big(m\ddot x_s+\gamma' \dot x_s+\partial_xV_{\alpha}(x_s)\Big)\right]\nonumber\\
&=\exp\{\Delta S_m[\{x_s\}]-\Delta S_{pu}\}
\end{align}
where  the subscripts $+$ and $-$ refer to the trajectories obtained with $\gamma'$ and $-\gamma'$, respectively. Choosing the appropriate backward path associated with a given forward path is always an issue in a nonequilibrium state. One has indeed the choice between changing or not changing the sign of the external parameters that specify the state, and the proper choice is the one that leads to a physically meaningful result for the concrete system under consideration\cite{TC2008}. This is the case here, but one must emphasize that changing $\gamma'$ into $-\gamma'$ is not a benign transformation: thermal fluctuations are then enhanced instead of being damped, and the Langevin dynamics does not lead to a stationary state at constant $\alpha$ if the effective friction coefficient $\gamma-\gamma'$ is negative. Although an equation similar to Eq. (\ref{Eqratio}) was derived in Ref. \cite{KQ2007} for a general velocity-dependent force $g(v)$,  this important issue was not reported and emphasis was only put on the additional entropy pumping contribution $(\gamma'/m) t=-\Delta S_{pu}$ in the exponential factor (on the other hand, the time-reversal symmetry of the control force is discussed in the previous Ref. \cite{KQ2004}). It turns out however that changing the sign of $\gamma'$  has also a significant consequence for the detailed FT, as  discussed below. Hereafter, the stochastic process (dynamics) with $\gamma'$ replaced by $-\gamma'$ is called the ``backward" process (dynamics) for brevity\cite{note1}. 

Starting from Eq. (\ref{Eqratio}), we now consider the ratio
\begin{align}
\label{EqR}
R[\{x_s\};p_0,p_1]\equiv \ln \frac{{\cal P}_+[\{x_s\}\vert x_0,v_0]p_0(x_0,v_0)}{{\cal P}_-[ \{\hat x_s\}\vert \hat x_0,\hat v_0]p_1(\hat x_0,\hat v_0)}=\Delta S_m[\{x_s\}]+\frac{\gamma'}{m} t +\ln \frac{p_0(x_0,v_0)}{p_1(x_t,-v_t)} 
\end{align}
where the probability distributions $p_0(x_0,v_0)$ and $p_1(\hat x_0,\hat v_0)$ for the initial and final states are still arbitrary at this stage. From Eq.(\ref{EqR}), one readily obtains the integral fluctuation relation
\begin{align}
\label{EqIFR}
 \ll e^{-R[\{x_s\}]}\gg =1 \ ,
\end{align}
where $ \ll...\gg$ denotes a path integral average over all possible paths $\{x_s\}_{s\in[0,t]}$ from $x(0)=x_0,\dot x(0)=v_0$  to  $x(t)=x_t, \dot x(t)=v_t$, and is defined by
\begin{align}
 \ll{\cal A} [\{x_s\}]\gg &\equiv \int dx_0 dv_0 \int dx_t dv_t \int_{(x_0,v_0)}^{(x_t,v_t)}{\cal D}x_s{\cal A}[\{x_s\}] {\cal P}_+[\{x_s\}\vert x_0,v_0]p_0(x_0,v_0)
 \end{align}
for any trajectory-dependent functional ${\cal A}[\{x_s\}]$. As usual, one must make a suitable choice of the `boundary' terms in Eq. (\ref{EqR}) (those that only depend on the distributions of the initial and final states) to give a physical interpretation to the functional $R[\{x_s\}]$\cite{S2005}.  The choice $p_1(x_t,v_t)=p(x,v,t)$, where $p(x,v,t)$  is the solution  of the Kramers equation for the given initial distribution $p_0(x_0,v_0)$, leads to
\begin{align}
R[\{x_s\}]=\Delta S_m[\{x_s\}]+\Delta S+\frac{\gamma'}{m} t \equiv \Sigma[\{x_s\}] \ ,
\end{align}
which is the total entropy production in the super-system along the specific trajectory $\{x_s\}_{s\in[0,t]}$. Then Eq. (\ref{EqIFR}) yields the integral fluctuation theorem (IFT)
\begin{align}
\label{EqIFT}
 \ll e^{-\Sigma[\{x_s\}]}\gg =1 
\end{align}
already given in Ref. \cite{KQ2007}. For a steady state at constant $\alpha$  characterized by the probability distribution
\begin{align}
\label{Eqphi}
p_{ss;\alpha}(x,v)\equiv \exp[-\phi_{\alpha}(x,v)] \ ,
\end{align}
the choice $p_0(x,v)=p_1(x,v)=p_{ss;\alpha}(x,v)$ yields $R[\{x_s\}]\equiv \Sigma_{ss,\alpha}[\{x_s\}]$, which from Eqs. (\ref{EqSm})-(\ref{EqSigma}) is given by
\begin{align}
\label{EqSigmass}
 \Sigma_{ss,\alpha}[\{x_s\}]=-\frac{\Delta E_{\alpha}}{T}+\Delta \phi_{\alpha}-\gamma' \big(\frac{1}{T}\int_0^t ds \: \dot x_s^2-\frac{t}{m}\big)
\end{align}
where $\Delta E_{\alpha}=(m/2)(v_t^2-v_0^2)+[V_{\alpha}(x_t)-V_{\alpha}(x_0)]$ and $\Delta \phi_{\alpha}\equiv \Delta S_{\alpha}=\phi_{\alpha}(x_t,v_t)-\phi_{\alpha}(x_0,v_0)$. In particular, the mean entropy production rate is
\begin{align}
\frac{\ll\Sigma_{ss,\alpha}[\{x_s\}]\gg}{t}=\frac{\gamma'}{m} \frac{T-T_{eff}}{T}
\end{align}
where $T_{eff}\equiv m<v^2>_{\alpha}$ (since $T_{eff}<T$ for $\gamma'>0$, the mean entropy production in the  super-system is thus a positive quantity, as it must be).
 
To derive the  stronger detailed fluctuation theorem for the entropy production in a NESS, we now consider the probability that  the functional $\Sigma_{ss,\alpha}[\{x_s\}]$ given by Eq. (\ref{EqSigmass}) takes a specific value $\Sigma=\sigma t$ along the forward trajectory (to simplify the notation, the subscript $\alpha$ is dropped hereafter). This probability is defined by 
\begin{align}
P_+(\Sigma_{ss}[\{x_s\}])=\sigma t)&\equiv  \ll\delta( \Sigma_{ss}[\{x_s\}]- \sigma t)\gg \ .
\end{align}
Using Eq.(\ref{EqR}), we then find
\begin{align}
P_+(\Sigma_{ss}[\{x_s\}]=\sigma t)&=\int dx_0 dv_0 \int dx_t dv_t \int_{(x_0,v_0)}^{(x_t,v_t)}{\cal D}x_se^{\Sigma[\{x_s\}]} {\cal P}_-[\{\hat x_s\}\vert x_t,v_t]p_{ss}(x_t,v_t)\delta( \Sigma_{ss}[\{x_s\}]- \sigma t)\nonumber\\
&=e^{\sigma t}\int dx_0 dv_0 \int dx_t dv_t \int_{(x_0,v_0)}^{(x_t,v_t)}{\cal D}x_s {\cal P}_-[\{\hat x_s\}\vert x_t,v_t]p_{ss}(x_t,v_t)\delta( \Sigma_{ss}[\{x_s\}]- \sigma t)\nonumber\\
&=e^{\sigma t}\int dx_0 dv_0 \int dx_t dv_t \int_{(x_0,v_0)}^{(x_t,v_t)}{\cal D}x_s {\cal P}_-[\{x_s\}\vert x_0,v_0]p_{ss}(x_0,v_0)\delta( \Sigma_{ss}[\{\hat x_s\}]- \sigma t)
\end{align}
where the integration variables $x_s$ and $\hat x_s$, $(x_0,v_0)$ and $(x_t,v_t)$ have been interchanged to obtain the last equality. This equation can be written in the form of a detailed FT as
\begin{align}
\label{EqDFT}
\frac{P_+(\Sigma_{ss}[\{x_s\}]=\sigma t)}{P_-(\hat\Sigma_{ss}[\{x_s\}]=-\sigma t)}=e^{\sigma t} \ ,
\end{align}
where the functional $\hat\Sigma_{ss}[\{x_s\}]$ is defined by
\begin{align}
\label{EqSigmasshat}
\hat\Sigma_{ss}[\{x_s\}]\equiv -\Sigma_{ss}[\{\hat x_s\}]=-\frac{\Delta E}{T}+\Delta \phi+\gamma' \big(\frac{1}{T}\int_0^t ds \: \dot x_s^2-\frac{t}{m}\big) \ ,
\end{align}
and $P_-(\hat\Sigma_{ss}[\{x_s\}]=-\Sigma )$ is the probability that  $\hat\Sigma_{ss}[\{x_s\}]$ takes the value $-\Sigma$ along a  trajectory in the backward process,  given that the initial state is sampled from the steady-state probability $p_{ss}(x,v)$ of the forward process. 

We thus see that the actual  FT (which is valid for any length $t$ of the trajectories) is more complicated that the one given in Ref. \cite{KQ2007}: the dynamics generating the stochastic trajectories in the numerator and the denominator are different, and so are the corresponding trajectory-dependent functionals (more precisely, the boundary terms $-\Delta E/T$ and $\Delta \phi$ in Eqs. (\ref{EqSigmass}) and (\ref{EqSigmasshat}) are identical whereas the sign of $\gamma'$ is changed in the  remaining `bulk' term).  Note also that $\hat\Sigma_{ss}[\{x_s\}]$ is {\it not} the entropy production functional in a steady state reached with the backward dynamics since the stationary distribution (and thus  $\Delta \phi $) is then different from the one given by Eq. (\ref{Eqphi}) (as an example, see  Eq. (\ref{Eqpss}) below). In fact, as already pointed out, there is no stationary distribution  with the backward dynamics if $\gamma'$ is larger than  the intrinsic friction $\gamma$ due to environment ($\gamma'>\gamma$ is the current situation in a cold damping setup since the goal is to reduce the thermal noise as much as possible\cite{LMC2000,JTCC2007,LKR2011}). Nevertheless, the FT given by Eq. (\ref{EqDFT}) is also valid in this case, as illustrated in Fig. 1 which shows the results of a numerical simulation of the Langevin equation for the harmonic potential studied in the next section. Note that the probability distributions for the forward and backward processes are quite different and that the latter (corresponding to a negative effective friction coefficient $\gamma-\gamma'$) exhibits a long tail on the positive side. However, the relation $P_+(\Sigma_{ss}[\{x_s\}]=\sigma t )=P_-(\hat\Sigma_{ss}[\{x_s\}]=-\sigma t)\:e^{\sigma t}$ is very well satisfied within the numerical accuracy of the calculation.
 
 \begin{figure}[hbt]
\begin{center}
\includegraphics[width=10cm]{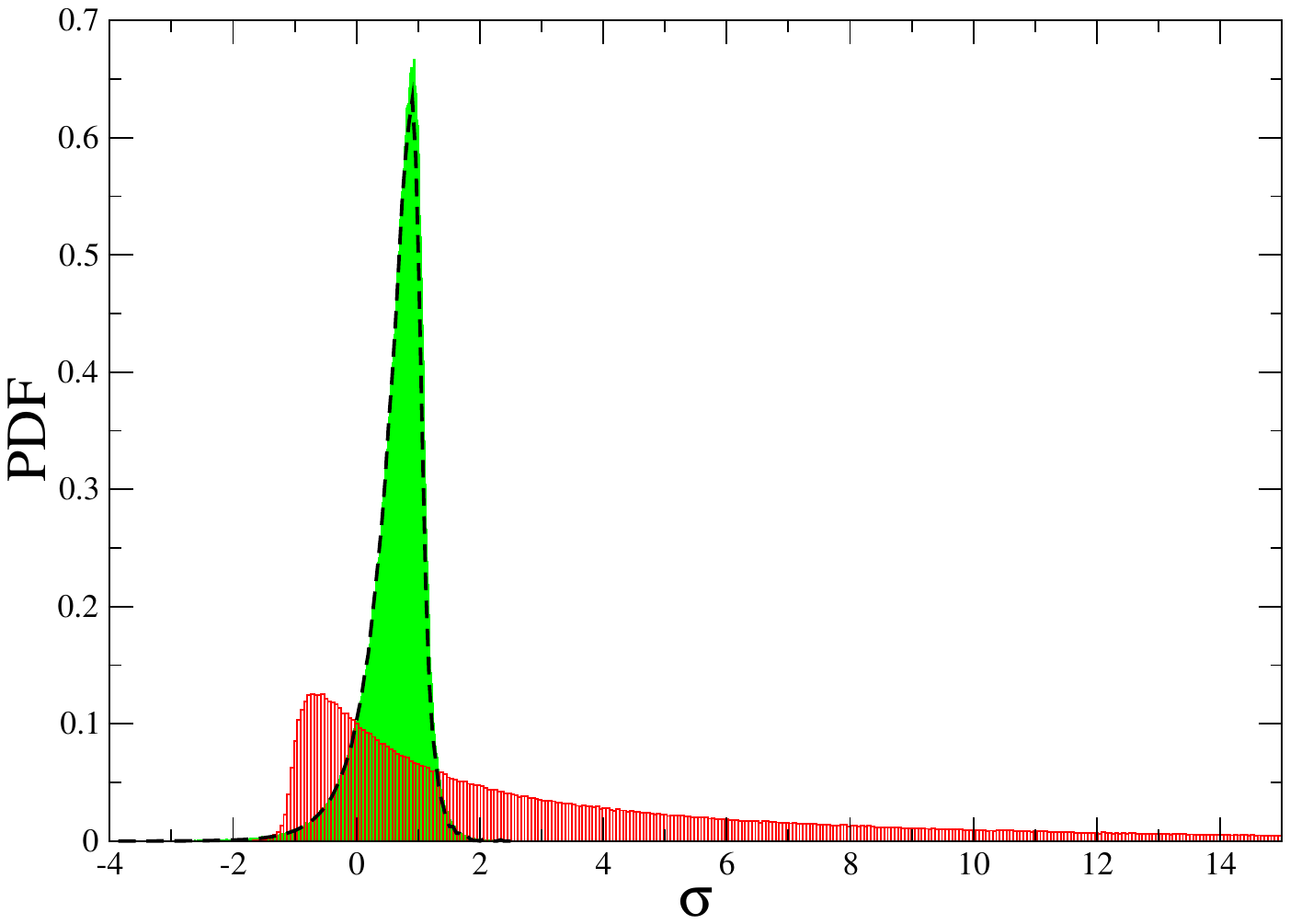}
 \caption{\label{Fig1} (Color online) Check of the detailed fluctuation theorem, Eq. (\ref{EqDFT}), for a stochastic harmonic oscillator with viscous dissipation and a velocity-dependent feedback force $g(v)=-\gamma' v$.  The figure shows the probability distribution functions $P_+(\Sigma_{ss}[\{x_s\}]=\sigma t )$ (green histogram) and $P_-(\hat\Sigma_{ss}[\{x_s\}]=\sigma t)$  (red histogram) for the forward and backward  processes, respectively. The black dashed line represents the product $P_-(\hat\Sigma_{ss}[\{x_s\}]=-\sigma t)\:e^{\sigma t}$ obtained from the histogram. The model parameters in Eq. (\ref{EqL2}) are  $k=0.2$,  $\gamma=1$, $\gamma'=1.2$ (with $m=1$ and $T=1$), and the observation time is $t=2$. The Langevin equation has been solved for $10^6$ realizations of the noise using Heun's method\cite{M2002} with  a time-step $\Delta t=0.001$.}
\end{center}
\end{figure}

It is clear that  the complexity of the FT for  a finite observation time $t$ comes from the fact that the boundary and bulk terms in $\Sigma_{ss}[\{x_s\}]$ do not behave in the same way under time reversal and/or reversal of the feedback force.  Therefore, one may expect some simplification if the contribution of the boundary term becomes negligible, which may occur in the long-time limit.
This supposes, however, that $\gamma'<\gamma$ so that a steady state can be reached asymptotically in the backward process. Then $\hat\Sigma_{ss}[\{x_s\}]$ becomes the actual entropy production in the steady state. This is illustrated in the next section by exact analytical calculations  for the harmonic oscillator.

 \section{Illustration on the stochastic harmonic oscillator}
 
\subsection{Entropy production probability distribution in the steady state}
 
To illustrate the preceding discussion, we now consider the paradigmatic case of a harmonic potential and we calculate the probability distribution function (PDF) of the entropy production in a steady state. The stochastic harmonic oscillator with viscous dissipation is relevant to the dynamics of an AFM cantilever\cite{PB2009},  to the motion of a colloidal particle in an optical trap, and to many other practical applications or nano-electromechanical systems  (see in particular Refs.\cite{V2008,B2009} for a recent application to the gravitational wave detector AURIGA). In general, it also permits a fully analytical analysis\cite{JGC2007}. 
In this case, Eq. (\ref{EqL1})  takes the very simple form
\begin{align}
\label{EqL2}
m\dot v_s =-kx_s-(\gamma+\gamma')v_s+\xi_s
\end{align}
where $k$ is the stiffness associated to the elastic force. The corresponding Kramers equation then writes 
\begin{align}
\partial_t p(x,v,t)=-v\partial_x p(x,v,t) +\frac{1}{m} \partial_v\left\{[kx+(\gamma +\gamma')v] p(x,v,t)+ \gamma \frac{T}{m} \partial_v p(x,v,t)\right\} \ ,
\end{align}
which has the stationary solution
\begin{align}
\label{Eqpss}
p_{ss}(x,v)=\frac{\sqrt{km}}{2\pi T}\frac{\gamma+\gamma'}{\gamma} \exp \{ -\frac{\gamma+\gamma'}{2\gamma T} [kx^2+mv^2]\} \ , 
\end{align} 
showing that the kinetic  temperature of the Brownian system is $T_{eff}=\gamma/(\gamma+\gamma') T $.
From Eq. (\ref{EqSigmass}), the total entropy production functional in the super-system is  given by
\begin{align}
\label{EqSharm}
\Sigma_{ss}[\{x_s\}]=\Sigma_{ss}^{(1)}(x_0,v_0,x_t,v_t)+\Sigma_{ss}^{(2)}[\{x_s\}]
\end{align}
where
\begin{align}
\label{EqSharm1}
\Sigma_{ss}^{(1)}(x_0,v_0,x_t,v_t)&=\frac{\gamma'}{2\gamma T}[k(x_t^2-x_0^2)+m(v_t^2-v_0^2)]
\end{align}
and 
\begin{align}
\label{EqSharm2}
\Sigma_{ss}^{(2)}[\{x_s\}]&= \frac{\gamma' }{m}\big(t-\frac{m}{T}\int_0^t ds \: \dot x^2_s\big)
\end{align}
are the boundary and bulk contributions, respectively (note that  both contributions vanish when $\gamma'=0$ since there is no other external force acting on the system which is then at equilibrium). $\Sigma_{ss}[\{x_s\}]$ is a quadratic functional of the noise and therefore its PDF is not Gaussian. Nevertheless, the generating or characteristic function defined by 
\begin{align}
\label{Eqcharac}
Z_{+}(\lambda,t)\equiv \ll e^{-\lambda \Sigma_{ss}[\{x_s\}] }\gg
\end{align}
can be explicitly calculated and the PDF is then recovered by taking the inverse Fourier transform
\begin{align}
\label{EqFourier}
P_{+}(\Sigma_{ss}[\{x_s\}]=\sigma t)=\frac{1}{2\pi i} \int_{-i\infty}^{+i\infty} d\lambda \: Z_{+}(\lambda,t)e^{\lambda \sigma t} 
\end{align}
where the integration is performed along the imaginary axis. We thus begin by computing $Z_{+}(\lambda,t)$. 

Inserting Eq. (\ref{EqSharm}) into Eq. (\ref{Eqcharac}), we obtain
\begin{align}
Z_{+}(\lambda,t)=e^{- \frac{\lambda \gamma' t}{m} }\int dx_0 dv_0 p_{ss}(x_0,v_0)\int dx_t dv_t  e^{-\frac{\lambda \gamma'}{2\gamma T }[k(x_t^2-x_0^2)+m(v_t^2-v_0^2)] }\int_{(x_0,v_0)}^{(x_t,v_t)} {\cal D}x_s \: {\cal P}_{+}[\{x_s\}\vert x_0,v_0] e^{ \frac{\lambda \gamma'}{T}\int_0^t ds\: \dot x^2_s}
\end{align}
where ${\cal P}_{+}[\{x_s\}\vert x_0,v_0] $ is given by Eq. (\ref{EqProba}).  This defines a new Lagrangian function 
\begin{align}
{\cal L}_{\lambda}(\ddot x_s,\dot x_s,x_s)&\equiv-\frac{1}{4\gamma T}(m\ddot x_s+(\gamma +\gamma')\dot x_s+kx_s)^2 + \frac{\lambda  \gamma'}{T} \dot x^2_s \ ,
\end{align}
which can be rewritten as 
 \begin{align}
{\cal L}_{\lambda}(\ddot x_s,\dot x_s,x_s)=-\frac{1}{4\gamma T}(m\ddot x_s+\tilde\gamma (\lambda)\dot x_s+kx_s)^2 -\frac{\gamma+\gamma'-\tilde\gamma }{2\gamma T}(k\:x_s\dot x_s+m\:\dot x_s\ddot x_s) 
\end{align}
by introducing  the $\lambda$-dependent damping coefficient
\begin{align}
\label{Eqgtilde}
 \tilde\gamma (\lambda) \equiv  [(\gamma+\gamma')^2-4\lambda \gamma \gamma']^{1/2} \ .
 \end{align}
This yields
\begin{align}
{\cal P}_{+}[\{x_s\}\vert x_0,v_0]e^{ \frac{\lambda \gamma'}{T}\int_0^t ds\: \dot x^2(s)}=\exp \left[\frac{\gamma + \gamma' -\tilde \gamma }{2m}t- \frac{\gamma +\gamma'-\tilde \gamma }{4\gamma T}[k(x_t^2-x_0^2)+m(v_t^2-v_0^2)]\right] \: {\cal P}_{\tilde \gamma }[\{x_s\}\vert x_0,v_0] 
\end{align}
where 
\begin{align}
\label{EqPtilde}
{\cal P}_{\tilde \gamma }[\{x_s\}\vert x_0,v_0] \equiv C \:\exp \left[\frac{\tilde \gamma }{2m}t-\frac{1}{4\gamma T}\int_0^t ds\: [m\ddot x_s+\tilde \gamma \dot x_s+kx_s]^2 \right] \ .
\end{align}
Hence
\begin{align}
\label{EqZ}
Z_{+}(\lambda,t)&= e^{\frac{\gamma + \gamma' -\tilde \gamma -2\lambda \gamma' }{2m} t}\int dx_0 dv_0 p_{ss}(x_0,v_0)\int dx_t dv_t  \exp\left[- \frac{\gamma +\gamma'-\tilde \gamma +2\lambda \gamma'}{4\gamma T}[k(x_t^2-x_0^2)+m(v_t^2-v_0^2)]\right]\nonumber\\
&\times \int_{(x_0,\dot x_0)}^{(x_t,\dot x_t)} {\cal D}[x(s)] \: {\cal P}_{\tilde \gamma}[\{x_s\}\vert x_0,v_0] \nonumber\\
&=\: e^{\mu(\lambda) t}\int dx_0 dv_0 p_{ss}(x_0,v_0)\int dx_t dv_t  \exp\left[- \frac{2m\mu(\lambda) +4\lambda \gamma'}{4\gamma T}[k(x_t^2-x_0^2)+m(v_t^2-v_0^2)]\right]P_{\tilde \gamma}(x_t,v_t,t\vert x_0,v_0,0)
\end{align}
where $\mu(\lambda)$ is defined by
\begin{align}
\label{Eqmu}
\mu(\lambda)\equiv \frac{\gamma + \gamma' -\tilde \gamma (\lambda) -2\lambda \gamma'}{2m}\ ,
\end{align}
and $P_{\tilde \gamma}(x_t,v_t,t\vert x_0,v_0,0)$ is the transition probability (or propagator) corresponding to the damping coefficient $\tilde \gamma(\lambda)$. Note that $Z_{+}(\lambda,t)$ as defined in the second line of Eq. (\ref{EqZ}) is properly normalized. Indeed, since $\tilde \gamma (\lambda=0)=\gamma + \gamma'$, one has $\mu(0)=0$ and thus
\begin{align}
Z_{+}(0,t)\equiv  \ll 1\gg&= \int dx_0 dv_0 p_{ss}(x_0,v_0)\int dx_t dv_t P_{\gamma+\gamma'}(x_t,v_t,t\vert x_0,v_0,0)\nonumber\\
&=\int dx_0 dv_0 \int dx_t dv_t P_{\gamma+\gamma'}(x_t,v_t,t;x_0,v_0,0)=1 \ .
\end{align}

Since the Langevin equation, Eq. (\ref{EqL2}), is linear and the noise is Gaussian, all stationary probability distributions are multivariate Gaussian distributions, and the explicit expression of $P_{\tilde \gamma}(x_t,v_t,t; x_0,v_0,0)$  is given by
\begin{align}
\label{EqP22}
P_{\tilde \gamma}(x_t,v_t,t; x_0,v_0,0)=\frac{1}{4\pi^2(\mbox{det}{\bf \Phi}_t)^{1/2}}\exp\{-\frac{1}{2} {\bf B}^T {\bf \Phi}_t^{-1} {\bf B}\}
\end{align}
where ${\bf \Phi}_t$ is the  matrix of time-correlation functions in the steady state
\[
{\bf \Phi}_t=\left(
\begin{array}{cccc}
  \phi_{xx}(0)&\phi_{xv}(0) & \phi_{xx}(t) & \phi_{xv}(t) \\
  \phi_{vx}(0)&\phi_{vv}(0) & \phi_{vx}(t) & \phi_{vv}(t)    \\
  \phi_{xx}(t)&\phi_{vx}(t) & \phi_{xx}(0) & \phi_{xv}(0)   \\
  \phi_{xv}(t)&\phi_{vv}(t) & \phi_{xv}(0) & \phi_{vv}(0) 
\end{array}
\right)=\left(
\begin{array}{cccc}
   \phi_{xx}(0)&0 & \phi_{xx}(t) & \dot \phi_{xx}(t) \\
  0&\phi_{vv}(0) & \dot \phi_{xx}(-t) &- \ddot\phi_{xx}(t)    \\
  \phi_{xx}(t)&\dot \phi_{xx}(-t) & \phi_{xx}(0)& 0   \\
  \dot \phi_{xx}(t)&-\ddot\phi_{xx}(t) & 0 & \phi_{vv}(0)
\end{array}
\right) 
\]
and  ${\bf B}$ is the $4$-dimensional vector representing the initial and final conditions
\[
\label{EqvectorB}
{\bf B}\equiv\left(
\begin{array}{c}
  x_0   \\
  v_0   \\
  x_t \\
  v_t    
\end{array} 
\right) \ .
\]
Here $\phi_{xx}(t)$ is the time-correlation function associated with an underdamped Langevin dynamics with damping coefficient $\tilde \gamma (\lambda)$\cite{R1989}, 
\begin{align}
\phi_{xx}(t)=\frac{\gamma T}{m^2(\omega_+^2 -\omega_-^2)}[-\frac{e^{-\omega_+ \vert t\vert}}{\omega_+}+\frac{e^{-\omega_-\vert t \vert}}{\omega_-}]
\end{align}
where 
\begin{align}
\omega_{\pm}=\frac{\tilde \gamma \pm \sqrt{\tilde \gamma ^2-4km}}{2m} \ .
\end{align}
In particular, $\phi_{xx}(0)=\gamma T/(\tilde \gamma k)$ and $\phi_{vv}(0)=\gamma T/(\tilde \gamma m)$.

From Eq. (\ref{EqP22}) we then compute the propagator $P_{\tilde \gamma}(x_t,v_t,t\vert x_0,v_0,0)=P_{\tilde \gamma}(x_t,v_t,t; x_0,v_0,0)/p_{\tilde \gamma}(x_0,v_0)$ where $p_{\tilde \gamma}(x_0,v_0)$ is obtained by replacing $\gamma+\gamma'$  by $\tilde \gamma$ in Eq. (\ref{Eqpss}). Inserting into Eq. (\ref{EqZ}), we find 
\begin{align}
Z_{+}(\lambda,t)=\: \frac{1}{4\pi^2(\mbox{det}{\bf \Phi}_t)^{1/2}}\frac{\gamma+\gamma'}{\tilde\gamma}e^{\mu(\lambda) t}\int dx_0 dv_0 \int dx_t dv_t  \exp\{-\frac{1}{2} {\bf B}^T {\bf (\Phi}_t^{-1}+{\bf L}) {\bf B}\}
\end{align}
where 
\[
{\bf L}=\frac{m}{\gamma T}\left(
\begin{array}{cccc}
 k \mu(\lambda)&0 &0 & 0 \\
  0&m \mu(\lambda)& 0 & 0  \\
 0&0 & k (\mu(\lambda)+\frac{2}{m}\lambda \gamma' )&0   \\
 0&0 &0& m (\mu(\lambda)+\frac{2}{m}\lambda \gamma' )
\end{array}
\right) \ .
\]
Carrying out the Gaussian integrals over $x_0,v_0$ and $x_t,v_t$, we finally obtain the compact expression
\begin{align}
\label{EqZplus}
Z_{+}(\lambda,t)=\: \frac{1}{[{\mbox{det}({\bf 1}+{\bf \Phi}_t{\bf L})]^{1/2}}}\frac{\gamma+\gamma'}{\tilde\gamma(\lambda)}e^{t \mu (\lambda)} 
\end{align}
which is the main result of this section.

\begin{figure}[hbt]
\begin{center}
\includegraphics[width=10cm]{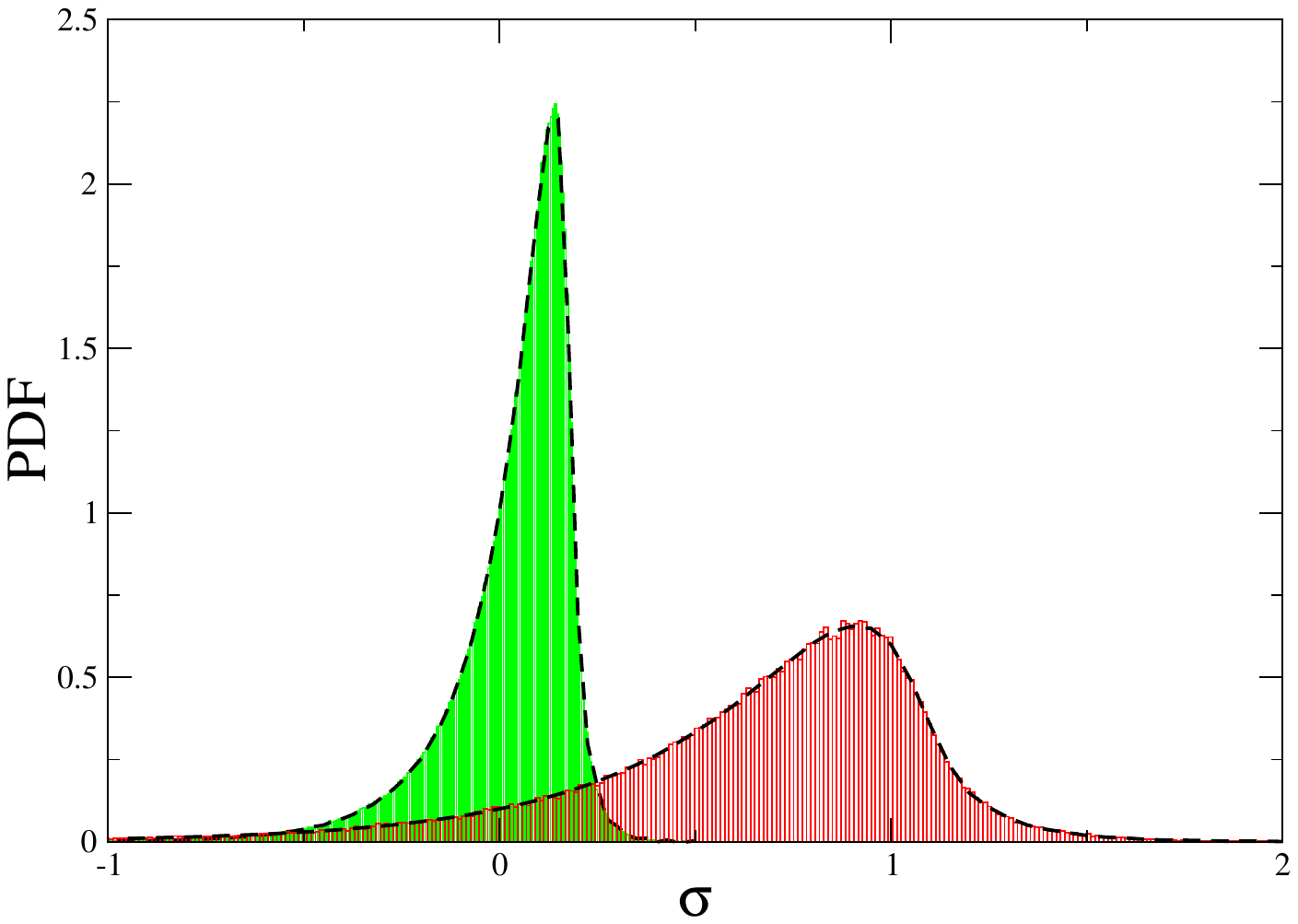}
 \caption{\label{FigProba} (Color online) Probability distribution $P_+(\Sigma_{ss}[\{x_s\}]=\sigma t)$ for $k=0.2$, $\gamma=1$, $\gamma'=0.2$ (green histogram) and $\gamma'=1.2$ (red histogram). The observation time is $t=2$. The numerical inverse Fourier transforms of Eq. (\ref{EqZplus}) (shown as dashed black lines) are compared to the histograms of $\Sigma_{ss}[\{x_s\}]$ obtained from the simulation of the Langevin equation.}
\end{center}
\end{figure}

This generating function is a complicated function of $\lambda$ and its inverse  Fourier transform can only be performed numerically. Note that  it was  implicitly assumed in the above calculations that the  damping coefficient $\tilde \gamma(\lambda)$ is real. From Eq. (\ref{Eqgtilde}),  this implies that $\lambda< \lambda_{max}=(\gamma+\gamma')^2/(4\gamma \gamma')$ on the real axis. On the other hand, the integration in  Eq. (\ref{EqFourier}) is performed along the imaginary axis and therefore quantities like $P_{\tilde \gamma}[\{x_s\}\vert x_0,v_0]$ (defined by Eq. (\ref{EqPtilde})) become complex and loose their physical meaning.  This may cast some doubt on the validity of the final result, Eq. (\ref{EqZplus}). However, one can show that the calculation remains valid  and that the inverse Fourier transform of Eq. (\ref{EqZplus})  is indeed a real quantity which correctly defines  the probability distribution $P_+(\Sigma[\{x_s\}]=\sigma t)$ (this will be clear below when considering the asymptotic long-time behavior). As shown in Fig. \ref{FigProba}, the numerical Fourier transform of the theoretical expression is indeed in excellent agreement with the histograms of $\Sigma_+[\{x_s\}]$ obtained from the direct simulation of the stochastic process,  for both $\gamma'<\gamma$ and $\gamma'>\gamma$.  One can  also clearly observe that the distributions are non-Gaussian.

Since $\tilde \gamma(0)=\gamma+\gamma'$ and $\mu(0)=0$ (which implies ${\bf L}={\bf 0}$), one readily sees  from Eq. (\ref{EqZplus}) that $Z_{+}(0,t)$  is properly normalized.  On the other hand, it is not immediately apparent  that the integral fluctuation theorem $Z_{+}(1,t)=1$ is satisfied. One needs to distinguish the two cases $\gamma>\gamma'$ and $\gamma<\gamma'$. In the first case, one has $\tilde \gamma(1)=\gamma-\gamma'$ so that $\mu(1)=0$. The calculation of the determinant in Eq. (\ref{EqZplus}) then gives
\begin{align}
\sqrt{\mbox{det}({\bf 1}+{\bf \Phi}_t{\bf L})}= \frac{\gamma+\gamma'}{\gamma-\gamma'} 
\end{align}
and thus  $Z_{+}(1,t)=1$, as it must be. Note that this result is obtained without using the explicit expressions of the time-correlation functions, but only their  values at $t=0$. In the second case, one has $\tilde \gamma(1)=\gamma'-\gamma$  and $\mu(1)=(\gamma-\gamma')/m$, and the calculation of the determinant gives
\begin{align}
\label{Eqdeter2}
\sqrt{\mbox{det}({\bf 1}+{\bf \Phi}_t{\bf L})}&= \frac{k m(\gamma'^2 -\gamma^2)}{\gamma^2 T^2}  [\phi_{xx}(t)\phi_{vv}(t)-\phi_{xv}(t)\phi_{vx}(t)] \nonumber\\
& =\frac{k}{m^3} \frac{\gamma'^2-\gamma^2}{(\omega_++\omega_-)^2\omega_+\omega_-}e^{-(\omega_++\omega_-)t}\nonumber\\
&=\frac{\gamma +\gamma'}{\gamma' -\gamma}e^{\frac{\gamma-\gamma'}{m}t} \ .
\end{align}
Inserting into Eq. (\ref{EqZplus}) yields  the correct result $Z_{+}(1,t)=1$. Remarkably, in this case, we had to use the explicit expressions of the time-correlation functions. 

In order to check the detailed FT expressed by Eq. (\ref{EqDFT}), one needs to calculate the generating function $Z_{-}(\lambda,t)$ of the functional $\hat\Sigma_{ss}[\{x_s\}]$ in the backward process. Formally, one can follow the same steps as in the preceding  calculation, at least up to Eq. (\ref{EqZ}) (replacing $\gamma'$  by $-\gamma'$). To proceed further, however, one needs to compute $P_{\tilde \gamma}(x_t,v_t,t; x_0,v_0,0)$, that is to solve the Kramers equation for the {\it backward} process with $p_{ss}(x_0,v_0)$ as initial condition (that is with the stationary distribution of the {\it forward} process). This is a complicated calculation which we have not performed (see Fig. 1 for a numerical check of Eq. (\ref{EqDFT})),  and in the following we shall only consider the asymptotic long-time regime. We just note that if the conventional FT were to hold exactly one would have 
\begin{align}
Z_{+}(1-\lambda,t)=Z_{-}(\lambda,t) \ .
\end{align}
From Eq. (\ref{Eqgtilde}), one  has $\tilde \gamma_{+} (1-\lambda)=\tilde \gamma_{-}(\lambda)$ so that $\mu_{+}(1-\lambda)=\mu_{-}(\lambda)$ where the indices $+$ and $-$ refer to $\gamma'$ and $-\gamma'$, respectively. Therefore one also has ${\bf \Phi}_{t,+}(1-\lambda)={\bf \Phi}_{t,-}(\lambda)$, assuming that the backward process has reached a steady-state (which implies that $\gamma'<\gamma$). The only function of $\lambda$ that does not have a simple symmetry is the matrix ${\bf L}$, which is not surprising since it contains the information about the initial conditions.

\subsection{Long-time behavior and large deviation function}

To complete this study we now consider the behavior of $P_+(\Sigma_{ss}[\{x_s\}]=\sigma t)$ for $t$ much larger than the effective viscous relaxation time  $\tau_r=m/(\gamma +\gamma')$. We expect  a large deviation form \cite{T2009} 
 \begin{align}
P_+(\Sigma_{ss}[\{x_s\}]=\sigma t)\sim e^{\:h(\sigma)t} 
\end{align}
where $h(\sigma)$ is the large deviation function (LDF) defined by 
\begin{align}
h(\sigma)=\lim_{t \rightarrow \infty} \frac{1}{t}\ln P_+(\Sigma_{ss}[\{x_s\}]=\sigma t)  \ .
\end{align}

As usual, the large-$t$ behavior of the PDF can be extracted from the integral representation (\ref{EqFourier}) by using a saddle-point approximation and taking care of the possible presence of singularities in the integrand. Hence we first need to derive  the asymptotic form of $Z_{+}(\lambda,t)$ from Eq. (\ref{EqZplus}), which is easily done by observing that the real parts of $\omega_{\pm}$ are always positive if $\tilde \gamma(\lambda)$ is real, that is if $\lambda<\lambda_{max}=(\gamma+\gamma')^2/(4\gamma \gamma')$. The time-correlation functions $\phi_{xx}(t),\phi_{xv}(t), \phi_{vv}(t)$ then go to zero as $t\rightarrow \infty$  and the matrix ${\bf \Phi}_t$ becomes diagonal.  The  determinant of ${\bf 1}+{\bf \Phi}_t{\bf L}$ in Eq. (\ref{EqZplus}) is readily calculated, and we obtain 
\begin{align}
\label{EqZasymp}
Z_{+}(\lambda,t)\sim g(\lambda) e^{t\mu(\lambda)} 
\end{align}
with
\begin{align}
\label{Eqg}
 g(\lambda) =\frac{4 (\gamma+\gamma')\tilde \gamma}{\vert (\gamma +\gamma' +\tilde \gamma)^2-4\lambda^2 \gamma'^2\vert} 
\end{align}
and $\mu(\lambda)$ defined by Eq. (\ref{Eqmu}). (Note that this calculation is not valid for $\lambda=1$ in the case $\gamma'>\gamma$ since then $\sqrt{\mbox{det}({\bf 1}+{\bf \Phi}_t{\bf L})}\rightarrow 0$, as can be seen from Eq. (\ref{Eqdeter2}). In this case $Z_{+}(1,t)=1$ for all $t$.)

By definition, $\mu(\lambda)= \lim_{t\rightarrow \infty} \ln Z(\lambda,t)/t$ is the cumulant generating function\cite{T2009} and the saddle point $\lambda^*(\sigma)$ is then solution of the equation
\begin{align}
\label{Eqsaddle}
\mu'(\lambda^*)+\sigma =0  
\end{align}
with 
\begin{align}
\mu'(\lambda)=\frac{\gamma'}{m} \frac{\gamma-\tilde \gamma(\lambda)}{\tilde \gamma(\lambda)}
\end{align}
from Eq. (\ref{Eqmu}). Since $\lim_{\lambda\rightarrow  -\infty}\mu'(\lambda)=-\gamma'/m$,  we see that the saddle point equation has no solution for $\sigma >\gamma'/m$. On the other hand, for $\sigma <\gamma'/m$, the solution of Eq. (\ref{Eqsaddle}) is given by
\begin{align}
\label{Eqlambdas}
\lambda^*(\sigma) = \frac{[\gamma'^2 -m\sigma  (\gamma+\gamma')][\gamma'^2 -m\sigma  (\gamma+\gamma')+2\gamma \gamma']}{4 \gamma \gamma' (\gamma' -m \sigma)^2}  \ ,
\end{align}
which is a function of $\sigma$ that  monotonically decreases from $\lambda_{max}$ to $-\infty$ as $\sigma$ increases from $-\infty$ to $\gamma'/m$ \cite{note2}. However, we also note from Eq. (\ref{Eqg}) that  the prefactor  $g(\lambda)$ diverges when $2\gamma'\lambda=\pm (\gamma +\gamma' +\tilde \gamma)$, and we thus have to determine for which values of $\sigma$ the saddle point coalesces with a pole of $g(\lambda)$.  To proceed further, we need to  consider the two cases $\gamma>\gamma'$ and $\gamma<\gamma'$ separately.

\subsubsection{$\gamma>\gamma'$}

In this case $g(\lambda)$ has a simple pole located on the real axis at $\lambda=\lambda_{min}\equiv-(2\gamma+\gamma')/\gamma'$ and by solving the equation $\lambda^*(\sigma)=\lambda_{min}$ we find that the saddle point hits this pole at $\sigma=\sigma^*$ with
\begin{align}
\label{Eqsigmastar}
\sigma^*=\frac{\gamma'}{m}\: \frac{2\gamma+\gamma'}{3\gamma +\gamma'} \ .
\end{align}
Starting from $\sigma= -\infty$, one can thus safely deform the contour of integration through the saddle point as long as $\sigma<\sigma^*$. The LDF  is then given by the Legendre transform of the cumulant generating function\cite{T2009}
\begin{align}
\label{Eqhsigma}
h(\sigma)=\mu(\lambda^*)+\lambda^* \sigma
\end{align}
which yields
\begin{align}
h(\sigma)\equiv h_1(\sigma)=-\frac{[\gamma'^2-m\sigma  (\gamma +\gamma')]^2}{4m\gamma \gamma'(\gamma'-m\sigma)}  \ .
\end{align}
On the other hand, for $\sigma>\sigma^*$, the steepest-descent contour must cross the pole and the leading contribution to the integral  comes from the pole (see the appendix for more details). The LDF is then a linear function of $\sigma$
\begin{align}
h(\sigma)=\mu(\lambda_{min})+\lambda_{min} \sigma 
\end{align}
which yields
\begin{align}
h(\sigma)\equiv h_2(\sigma)&=\frac{\gamma+\gamma'}{m}-\frac{2\gamma+\gamma'}{\gamma'} \sigma \ .
\end{align}

\begin{figure}[hbt]
\begin{center}
\includegraphics[width=10cm]{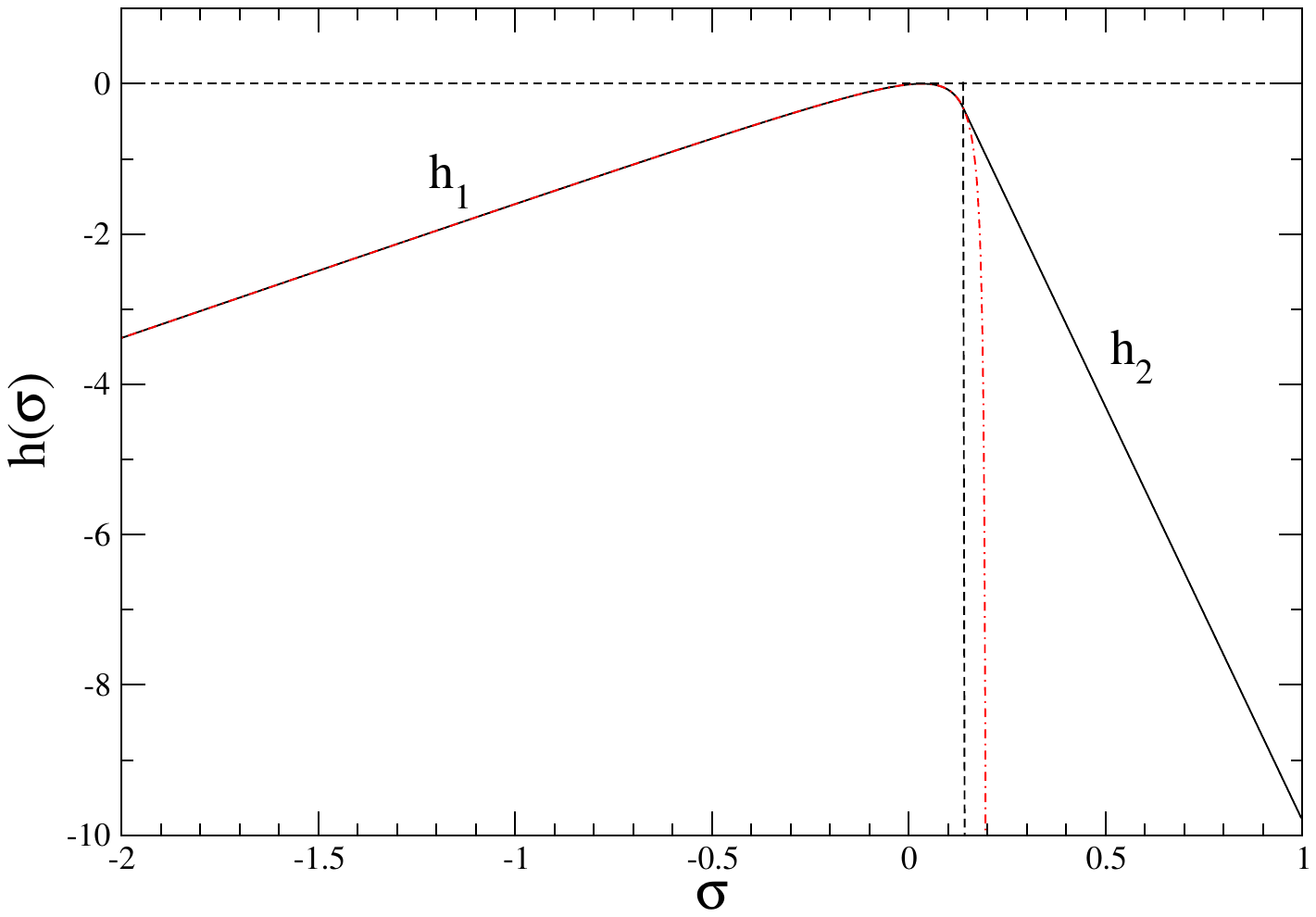}
 \caption{\label{FigLDF} (Color online) Large deviation function $h(\sigma)$ for $\gamma>\gamma'$ ($m=1,\gamma=1,\gamma'=0.2$). The vertical dashed line marks the position of $\sigma^*$ that separates the branches $h_1(\sigma)$ and $h_2(\sigma)$. The (red) dotted-dashed line represents the branch $h_1(\sigma)$ for $\sigma>\sigma^*$ which diverges  at $\sigma=\gamma'/m$. Note  that $h_1(\sigma)=0$ for $\sigma=\gamma'^2/[m(\gamma+\gamma')]$, the mean entropy production rate.}
\end{center}
\end{figure}

The behavior of the LDF as a function of $\sigma$ is illustrated numerically in Fig. \ref{FigLDF} (one can easily check that $h(\sigma)$ and its first derivative are continuous at $\sigma=\sigma^*$). In the appendix, we  give the complete asymptotic form of the probability distribution $P_+(\Sigma_{ss}[\{x_s\}]=\sigma t)$, taking into account the proximity of the saddle point to the pole as explained in Ref. \cite{W1989}, and the contribution of the residue of $g(\lambda)$ at $\lambda=\lambda_{min}$ when the pole is crossed.  Interestingly,  the PDF in the long-time limit becomes independent of $k$, the stiffness of the harmonic oscillator  (see Ref. \cite{S2011} for a similar observation). 

Before considering the case $\gamma<\gamma'$, let us briefly comment these results. We first note that the pole in $g(\lambda)$, which limits the position of the saddle point, appears when the average over the initial and final states is performed. Its presence can be traced back to the contribution of $\Sigma_{ss}^{(1)}$, the boundary term in $\Sigma_{ss}[\{x_s\}]$ given by Eq. (\ref{EqSharm1}). The singularity arises because the position and the velocity of the particle are unbounded and Gaussian distributed according to the steady-state probability distribution (\ref{Eqpss}). As a consequence, the PDF of  $\Sigma_{ss}^{(1)}$ has an exponentially decreasing tail (specifically, of the form $e^{-\frac{\gamma+\gamma'}{\gamma'} \vert \sigma \vert t}$) and large fluctuations of order $t$ may occur which cannot be neglected in the sum $\Sigma_{ss}^{(1)}+\Sigma_{ss}^{(2)}$ despite the fact that  $\Sigma_{ss}^{(1)}$ is not extensive in time. On the other hand, there is no singularity if the initial and final positions and velocities are  fixed. Such an interplay between boundary and bulk terms  is well documented in the literature on  large deviations\cite{T2009} and fluctuation relations (see e.g. Refs. \cite{F2002,VC2003,HS2007,BGGZ2006,V2006,PRV2006,MPRV2008} and more recently Ref. \cite{S2011}). In the present case, however, this interplay is a bit unusual. Indeed,  the slope of $h_2(\sigma)$ is not equal to $-(\gamma+\gamma')/\gamma'$, which means that it is {\it not} simply imposed by the exponential tail of the PDF of $\Sigma_{ss}^{(1)}$ (in contrast, for instance, with the model studied in Ref. \cite{VC2003}). Clearly, one cannot treat the fluctuations of the boundary and bulk contributions independently, even asymptotically.  This is all the more remarkable that the latter (divided by $t$) is bounded by $\gamma'/m$ as can be readily seen from Eq. (\ref{EqSharm2}). This is actually  the origin of the divergence in $\lambda^*(\sigma)$ and $h_1(\sigma)$ at  $\sigma=\gamma'/m$. However, since $\sigma^*<\gamma'/m$, this divergence occurs in the region where the large fluctuations are described by $h_2(\sigma)$ and it is thus harmless (see Fig. \ref{FigLDF}).

We can now come back to the detailed fluctuation theorem (Eq. (\ref{EqDFT})) by noting that the function $h_1(\sigma)$ possesses the symmetry
\begin{align}
\label{Eqsymh1}
h_1^+(\sigma)-h_1^-(-\sigma)= \sigma 
\end{align}
where the superscripts $+$ and $-$ refer to $\gamma'$ and $-\gamma'$ respectively (hence $h_1^+(\sigma)\equiv h_1(\sigma)$). This is precisely the asymptotic limit of the FT
\begin{align}
\lim_{t \rightarrow \infty}\frac{1}{t}\ln \frac{P_+(\Sigma_{ss}[\{x_s\}=\sigma t]}{P_-(\hat\Sigma_{ss}[\{x_s\}=-\sigma t]}=\sigma 
\end{align}
provided that the large deviation forms
\begin{subequations}
\begin{align}
\label{EqasymP}
P_+(\Sigma_{ss}[\{x_s\}]&=\sigma t)\sim e^{\:h_1^+(\sigma)t} \\
P_-(\hat \Sigma_{ss}[\{x_s\}]&=\sigma t)\sim e^{\:h_1^-(\sigma)t} 
\end{align}
\end{subequations}
are both valid. We already know that Eq. (63a) is valid for $\sigma<\sigma^*$.  Eq. (63b) is also valid when the fluctuations of the boundary term becomes negligible so that it is irrelevant to sample the initial state of the backward process with the steady-state probability of the forward process. Then Eq. (\ref{EqZasymp}) (with $\gamma'$ replaced by $-\gamma'$ ) correctly describes asymptotically  the generating function $Z_{-}(\lambda,t)$ of the functional $\hat\Sigma_{ss}[\{x_s\}]$. This occurs for 
$\sigma>\sigma_-^*\equiv \sigma^*(-\gamma')$. One can check from Eq. (\ref{Eqsigmastar}) that  $\sigma^*>-\sigma_-^*$ and therefore Eq. (\ref{Eqsymh1}) is indeed the asymptotic expression of the detailed FT  for $\sigma<-\sigma^*_-$ (in the case displayed in Fig. \ref{FigLDF}, $\sigma^*=0.1375$ and  $\sigma^*_-\approx -0.129$).

\subsubsection{$\gamma<\gamma'$}

We now turn to the physically more relevant case $\gamma<\gamma'$. The new feature is the presence of a second pole in $g(\lambda)$ at $\lambda=1$ (see also the remark after Eq. (\ref{Eqg})). By solving the equation $\lambda^*(\sigma)=1$, we find that the saddle point and the pole coalesce  at $\sigma=\sigma^{**}$ with
\begin{align}
\sigma^{**}= \frac{\gamma'}{m} \: \frac{\gamma'-2\gamma}{\gamma'-\gamma} \ .
\end{align}
This value is smaller than $\sigma^*$ and therefore  the LDF is  described by the function $h_1(\sigma)$ (the Legendre transform of the cumulant generating function)  in the interval $[\sigma^{**},\sigma^*]$ only. On the other hand, for $\sigma\le \sigma^{**}$, the leading contribution to the integral comes from the pole at $\lambda=1$ and the LDF is again linear
\begin{align}
h(\sigma)=\mu(1)+\sigma 
\end{align}
which yields
\begin{align}
h(\sigma)\equiv h_3(\sigma)&=-\frac{\gamma'-\gamma}{m}+ \sigma \ .
\end{align}
Finally, for  $\sigma \ge \sigma^*$, the contribution from the other pole  is dominant and $h(\sigma)\equiv h_2(\sigma)$ like in the case $\gamma>\gamma'$. 
 \begin{figure}[hbt]
\begin{center}
\includegraphics[width=9.5cm]{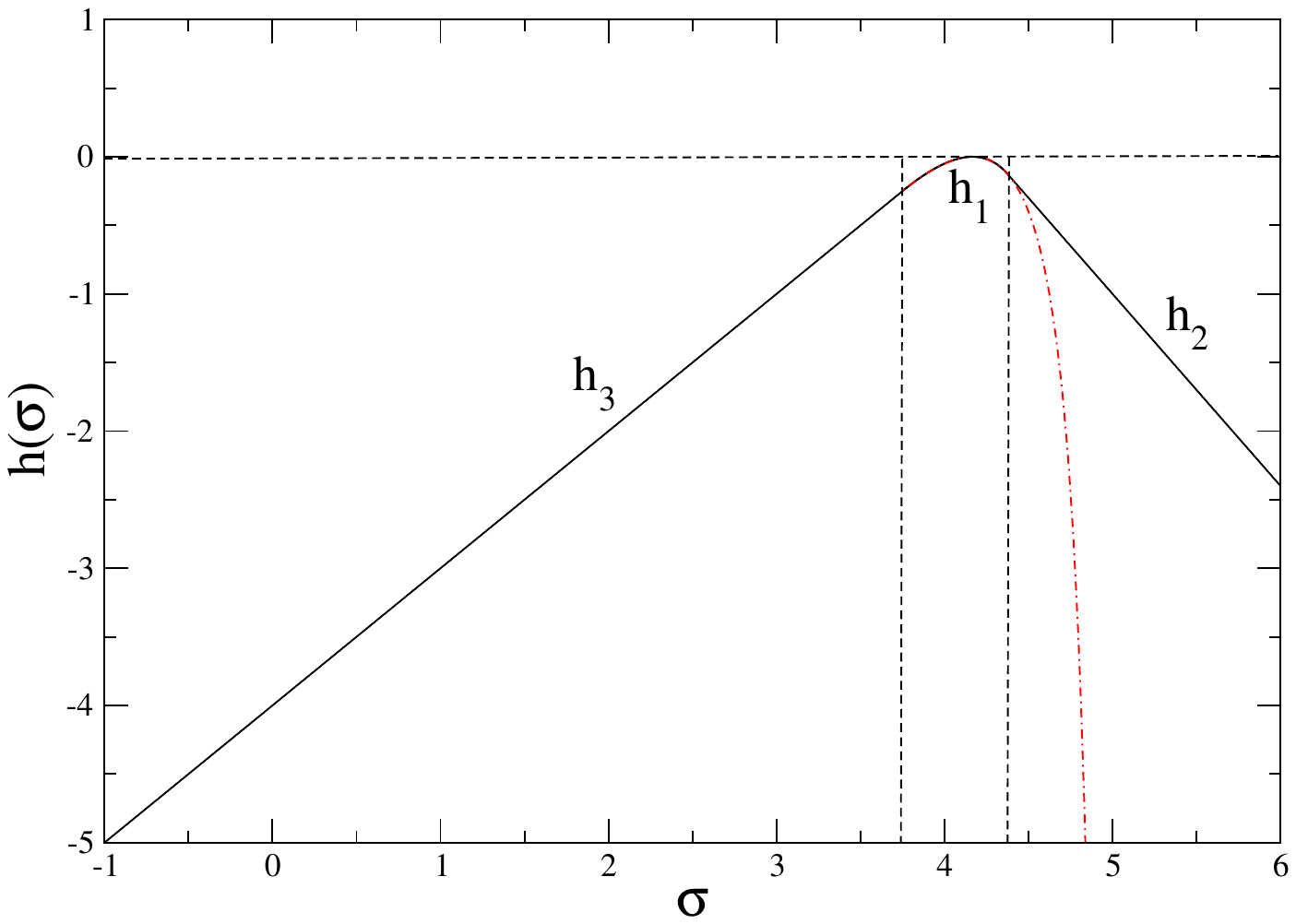}
 \caption{\label{FigLDF2} (Color online) Large deviation function $h(\sigma)$ for $\gamma<\gamma'$ ($m=1,\gamma=1,\gamma'=5$). The  vertical dashed lines mark the positions of $\sigma^{**}$ and $\sigma^*$ that separate the branches $h_3(\sigma)$, $h_1(\sigma)$, and $h_2(\sigma)$. The (red) dotted-dashed line represents the branch $h_1(\sigma)$ for $\sigma>\sigma^*$ which diverges at $\sigma=\gamma'/m$.}
\end{center}
\end{figure}

The behavior of the LDF as a function of $\sigma$ is illustrated numerically in Fig. \ref{FigLDF2} and the complete asymptotic expression of the PDF (taking into account the presence of the two poles in $g(\lambda)$) is also given in the appendix. Fig. \ref{Figasymp} in the appendix confirms that this expression correctly describes the PDF when the observation time $t$ is very large. 

Note that $\sigma^*\sim \sigma^{**} \sim \gamma'/m$ when $\gamma'/\gamma \gg 1$ so that the domain of validity of the central branch $h_1(\sigma)$ becomes very small. In any case,  the symmetry relation (\ref{Eqsymh1})  cannot be interpreted  as the asymptotic expression of the detailed FT when $\gamma'>\gamma$. Indeed, as already stressed, the system does not reach a steady state with the `backward' dynamics and Eq. (\ref{EqZplus}) does not describe the generating function $Z_{-}(\lambda,t)$ of the functional $\hat\Sigma_{ss}[\{x_s\}]$ (we recall that the expressions of the steady-state time-correlation functions have been used to derive this equation).  

Finally, let us stress that a similar analysis can be performed for the probability distribution functions of the heat adsorbed by the oscillator or the injected power (the expressions of the characteristic functions are quite similar to Eq. (\ref{EqZplus})). The analytical results can be directly compared to the data collected by the gravitational wave detector AURIGA  which are given in Ref.\cite{B2009}.

\section{Concluding remarks}

To summarize, we have revisited the model of a classical molecular refrigerator described by an underdamped Langevin equation with a feedback force proportional to the velocity.  Unlike the viscous force due to the environment, the feedback can be seen as a virtual viscous force that creates dissipation without introducing fluctuations. This modifies the entropy production in the system and the  contribution of the feedback mechanism to the entropy must be included in the second law and the fluctuation theorems, as discussed previously\cite{KQ2004,KQ2007}. However, we have shown that the detailed fluctuation theorem has a more complicated interpretation than originally suggested\cite{KQ2007}. This results from the fact that  the sign of $\gamma'$, the friction coefficient  associated to the feedback force, must be changed in order to determine the appropriate backward (time-reversed) path corresponding to a given forward  path in the path integral approach. This kind of issue has already been discussed in the literature\cite{TC2008} but it takes a special importance here due to the friction-like character of the feedback force\cite{KQ2004}. For instance, this implies that the system is heated and cannot reach a stationary state in the backward process if  $\gamma'$ is larger than $\gamma$, the intrinsic friction due to the environment. $\gamma'>\gamma$ is  in fact  the common experimental situation. By solving analytically the harmonic oscillator and computing the probability distribution of the total entropy production in a NESS  we have shown that the regime of fluctuations in the cooling mode (the usual forward process) also depends on whether the ratio $\gamma'/\gamma$  is smaller or larger than $1$. In particular, the large time behavior of the PDF, as described by the large deviation function, is  controlled by a subtle and rather unusual interplay between boundary and bulk contributions. This is a remarkable feature taking into account the simplicity of the model and it might be an interesting challenge to check this behavior experimentally.

\acknowledgments

We  thank H. Touchette for very helpful comments about the calculation and the interpretation of large deviations in the entropy production and for a critical reading of the manuscript. 

\appendix 

\section{Asymptotic expression of the PDF}

In this appendix we give the asymptotic form of the probability distribution $P_+(\Sigma_{ss}[\{x_s\}]=\sigma t)$ in the long-time limit. To take into account the proximity of the saddle point to a pole in $g(\lambda)$  (for instance, for $\gamma>\gamma'$, the pole at $\lambda_{min}$ which is reached when $\sigma=\sigma^*$), we write
\begin{align}
g(\lambda)=\frac{g_{-1}^*}{\lambda-\lambda_{min}} +\tilde g(\lambda)
\end{align}
where $g_{-1}^*$ is the residue of $g(\lambda)$ at $\lambda_{min}$ and  $\tilde g(\lambda)$ is the regular part. We then treat the two contributions to the contour integral as explained in Ref.\cite{W1989} and add the contribution of the residue when the contour has to cross the pole (see also Ref.\cite{S2011} for a  similar calculation). Similarly, in presence of the other pole at  $\lambda=1$ for $\gamma<\gamma'$, we write
\begin{align}
g(\lambda)=\frac{g_{-1}^*}{\lambda-\lambda_{min}} +\frac{g_{-1}^{**}}{\lambda-1} +\hat g(\lambda) \ .
\end{align}

Skipping the details, we find:

a) For $\gamma >\gamma'$,

\begin{equation}
\label{Eqasym1}
P_+(\Sigma_{ss}[\{x_s\}]=\sigma t)\approx \left \{\begin{aligned}
&\frac{e^{h_1(\sigma)t}}{\sqrt{\pi t}}f_1(\sigma)+\frac{1}{2}e^{h_2(\sigma)t}g_{-1}^*\mbox{erfc}\big(\sqrt{tu(\sigma)}\big) & \quad \mbox{for $\sigma \le \sigma^*$}\\       
&\frac{e^{h_1(\sigma)t}}{\sqrt{\pi t}}f_2(\sigma)+e^{h_2(\sigma)t}g_{-1}^*[1-\frac{1}{2}\mbox{erfc}\big(\sqrt{tu(\sigma)}\big)] & \quad \mbox{for $\sigma^*\le \sigma\le \frac{\gamma'}{m}$} \\
&g_{-1}^*e^{h_2(\sigma)t} & \quad \mbox{for $\sigma\ge \frac{\gamma'}{m}$}\\
\end{aligned}
 \right.
\end{equation}
where
\begin{align}
u(\sigma)&=h_2(\sigma)-h_1(\sigma)\ ,
\end{align}
\begin{align}
f_1(\sigma)&=\frac{g(\lambda^*(\sigma))}{\sqrt{2\mu''(\lambda^*(\sigma))}}-\frac{g_{-1}^*}{2\sqrt{u(\sigma)}} \nonumber\\
f_2(\sigma)&=\frac{g(\lambda^*(\sigma))}{\sqrt{2\mu''(\lambda^*(\sigma))}}+\frac{g_{-1}^*}{2\sqrt{u(\sigma)}} \ ,
\end{align}
and 
\begin{align}
g_{-1}^*&=\frac{(\gamma+\gamma')(3\gamma+\gamma')^2}{2\gamma'(2\gamma+\gamma')^2} \ .
\end{align}

b) For $\gamma <\gamma'$, 

\begin{figure}[hbt]
\begin{center}
\includegraphics[width=10cm]{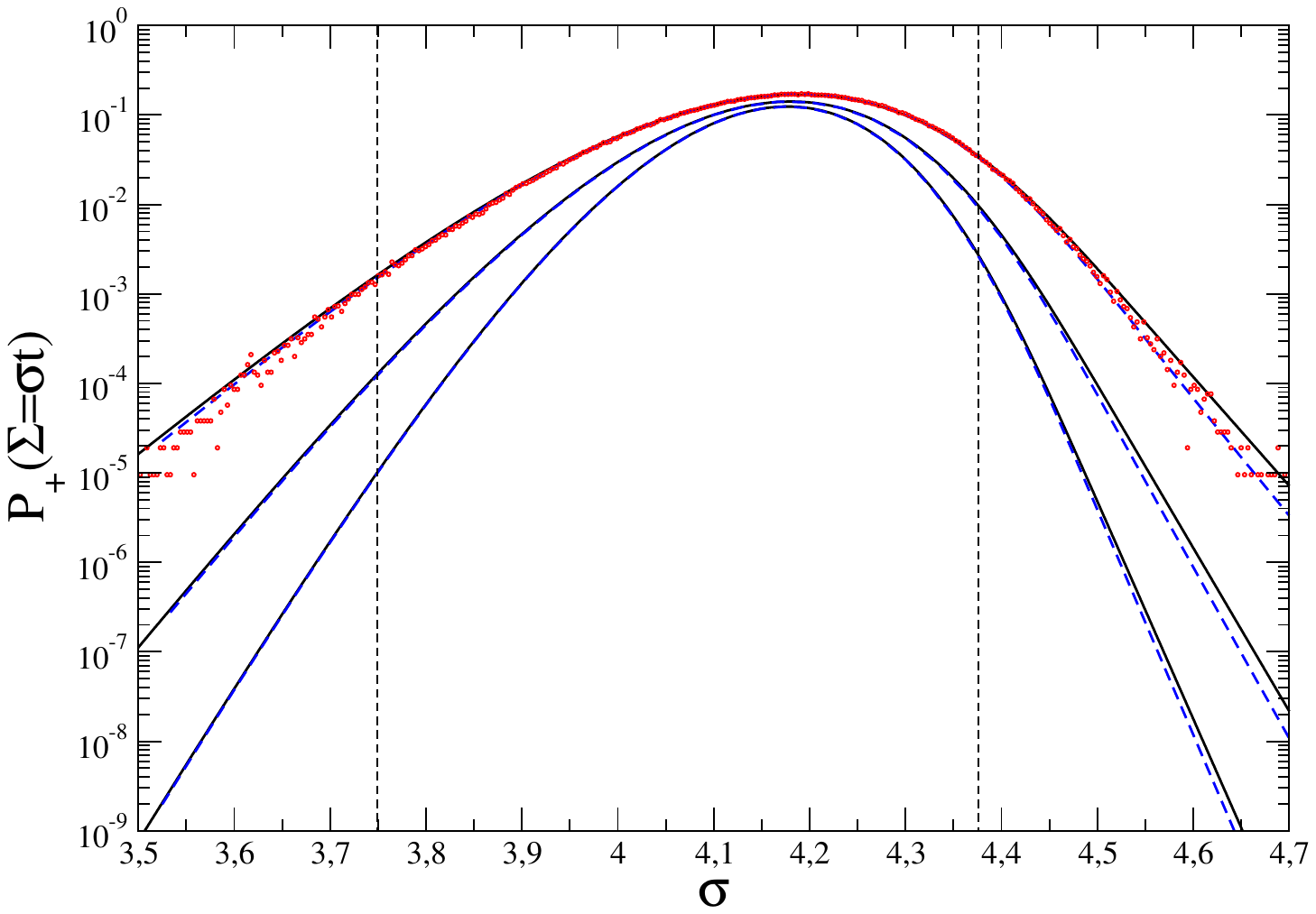}
 \caption{\label{Figasymp} (Color online) PDF of the entropy production for $\gamma=1$ and $\gamma'=5$ ($T=1,m=1,k=0.2$) and $t=20,30,40$ (from top to bottom). The solid black lines represent the analytical asymptotic expressions given by Eqs. (\ref{Eqasym2}) and the dashed blue lines are the corresponding numerical inverse Fourier transforms of $Z_{+}(\lambda,t)$. The red points are obtained from the numerical simulation of the Langevin equation for $t=20$. The vertical dashed lines mark the positions of $\sigma^*$ (right) and $\sigma^{**}$ (left).}
\end{center}
\end{figure}

\begin{equation}
\label{Eqasym2}
P_+(\Sigma_{ss}[\{x_s\}]=\sigma t)\approx \left \{\begin{aligned}
&\frac{e^{h_1(\sigma)t}}{\sqrt{\pi t}}f_1(\sigma)+\frac{1}{2}e^{h_2(\sigma)t}g_{-1}^*\mbox{erfc}\big(\sqrt{tu(\sigma)}\big) +e^{h_3(\sigma)t}g_{-1}^{**}[\frac{1}{2}\mbox{erfc}\big(\sqrt{tv(\sigma)}\big)-1] & \quad \mbox{for $\sigma \le \sigma^{**}$}\\
&\frac{e^{h_1(\sigma)t}}{\sqrt{\pi t}}f_2(\sigma)+\frac{1}{2}e^{h_2(\sigma)t}g_{-1}^*\mbox{erfc}\big(\sqrt{tu(\sigma)}\big)- \frac{1}{2}e^{h_3(\sigma)t}g_{-1}^{**}\mbox{erfc}\big(\sqrt{tv(\sigma)}\big)& \quad \mbox{for $\sigma^{**}\le \sigma\le \sigma^*$}\\
&\frac{e^{h_1(\sigma)t}}{\sqrt{\pi t}}f_3(\sigma)+e^{h_2(\sigma)t}g_{-1}^*[1-\frac{1}{2}\mbox{erfc}\big(\sqrt{tu(\sigma)}\big)]-\frac{1}{2}e^{h_3(\sigma)t}g_{-1}^{**}\mbox{erfc}\big(\sqrt{tv(\sigma)}\big)  & \quad \mbox{for $\sigma^*\le \sigma \le \frac{\gamma'}{m}$}\\
&g_{-1}^*e^{h_2(\sigma)t}  & \quad \mbox{for $ \sigma \ge\frac{\gamma'}{m}$}\\
\end{aligned}
 \right.
\end{equation}
where
\begin{align}
v(\sigma)=h_3(\sigma)-h_1(\sigma) \ ,
\end{align}
\begin{align}
f_1(\sigma)&=\frac{g(\lambda^*(\sigma))}{\sqrt{2\mu''(\lambda^*(\sigma))}}-\frac{g_{-1}^*}{2\sqrt{u(\sigma)}}-\frac{g_{-1}^{**}}{2\sqrt{v(\sigma)}} \nonumber\\
f_2(\sigma)&=\frac{g(\lambda^*(\sigma))}{2\sqrt{2\mu''(\lambda^*(\sigma))}}-\frac{g_{-1}^*}{\sqrt{u(\sigma)}}+\frac{g_{-1}^{**}}{2\sqrt{v(\sigma)}} \nonumber\\
f_3(\sigma)&=\frac{g(\lambda^*(\sigma))}{2\sqrt{2\mu''(\lambda^*(\sigma))}}+\frac{g_{-1}^*}{\sqrt{u(\sigma)}}+\frac{g_{-1}^{**}}{2\sqrt{v(\sigma)}}  \ ,
\end{align}
and 
\begin{align}
g_{-1}^{**}=-\frac{(\gamma'-\gamma)^2(\gamma'+\gamma)}{2\gamma'^3} \ .
\end{align}

As shown in Fig. \ref{Figasymp}, the above asymptotic expressions are in excellent agreement with the numerical inverse Fourier transform of $Z_{+}(\lambda,t)$ (Eq. (\ref{EqZplus})). In particular, we note that the small discrepancies  on the right hand side (for $\sigma>\sigma^*$) diminish as $t$ increases. For $t=20$, there is also a good agreement with the numerical simulation of the Langevin equation.


\begin{thebibliography}{10}
\bibitem{note2} The fact that the saddle point is smaller than $\lambda_{max}$ is not surprising. Indeed, the analytical continuation of the integral (\ref{EqFourier}) to real values of $\lambda$ must be real and non-negative to properly define a probability distribution. Since both $\mu(\lambda)$ and $g(\lambda)$ depend on  $\tilde \gamma (\lambda)$, this implies that $\lambda^*<\lambda_{max}$.
\end{thebibliography}
\end{document}